
\input phyzzx
\hoffset=1truein
\voffset=1.0truein
\hsize=6truein
\def\TITLEPAGE{\frontpagetrue}
\def\CALT#1{\hbox to\hsize{\tenpoint \baselineskip=12pt
	\hfil\vtop{\hbox{\strut CALT-68-#1}
	\hbox{\strut DOE RESEARCH AND}
	\hbox{\strut DEVELOPMENT REPORT}}}}

\def\CALTECH{\smallskip
	\address{California Institute of Technology, Pasadena, CA
91125}}

\def\AUTHOR#1{\vskip .5in \centerline{#1}}
\def\ANDAUTHOR#1{\smallskip \centerline{\it and} \smallskip
\centerline{#1}}
\def\ABSTRACT#1{\vskip .5in \vfil \centerline{\twelvepoint \bf
Abstract}
	#1 \vfil}
\def\ENDTITLEPAGE{\vfil\eject\pageno=1}

\def\sqr#1#2{{\vcenter{\hrule height.#2pt
      \hbox{\vrule width.#2pt height#1pt \kern#1pt
        \vrule width.#2pt}
      \hrule height.#2pt}}}

\def\section#1#2{
\noindent\hbox{\hbox{\bf #1}\hskip 10pt\vtop{\hsize=5in
\baselineskip=12pt \noindent \bf #2 \hfil}\hfil}
\medskip}

\def\underwig#1{	
	\setbox0=\hbox{\rm \strut}
	\hbox to 0pt{$#1$\hss} \lower \ht0 \hbox{\rm \char'176}}

\def\bunderwig#1{	
	\setbox0=\hbox{\rm \strut}
	\hbox to 1.5pt{$#1$\hss} \lower 12.8pt
	 \hbox{\seventeenrm \char'176}\hbox to 2pt{\hfil}}

\def\MEMO#1#2#3#4#5{
\frontpagetrue
\centerline{\tencp INTEROFFICE MEMORANDUM}
\smallskip
\centerline{\bf CALIFORNIA INSTITUTE OF TECHNOLOGY}
\bigskip
\vtop{\tenpoint
\hbox to\hsize{\strut \hbox to .75in{\caps to:\hfil}\hbox to
3.8in{#1\hfil}
\quad\the\date\hfil}
\hbox to\hsize{\strut \hbox to.75in{\caps from:\hfil}\hbox to
3.5in{#2\hfil}
\hbox{{\caps ext-}#3\qquad{\caps m.c.\quad}#4}\hfil}
\hbox{\hbox to.75in{\caps subject:\hfil}\vtop{\parindent=0pt
\hsize=3.5in #5\hfil}}
\hbox{\strut\hfil}}}
\tolerance=10000
\hfuzz=5pt
\def\dev{B\rightarrow De\bar \nu_e}
\def\deva{B\rightarrow D^*e \bar \nu_e}

\def\vslash{\rlap{/}v}
\TITLEPAGE
\CALT{1862}         

\bigskip
\titlestyle {Corrections from Low Momentum Physics to Heavy\break
Quark Symmetry

Relations for  $\dev$ and $\deva$ Decay

\foot{Work supported in part by the U.S. Dept. of Energy
under Grant No. DE-FG03-92-ER40701.}}
\AUTHOR{Chi-Keung Chow}
\ANDAUTHOR{Mark B. Wise}
\CALTECH
\ABSTRACT{Heavy quark symmetry relations for $\dev$ and $\deva$ form

factors receive
corrections suppressed by powers of the charm quark mass.  For small
up
and down quark massed the leading corrections of order
$(1/m_c)^{n+2}~n
= 0,1,2,...$ have a nonanalytic dependence on the light quark masses
of
the form $\ell n m_\pi^2$ for $n = 0$ and $(1/m_\pi)^n$ for $n = 1,2,
...$~.  These corrections arise from very low momentum physics (i.e.,
well below the chiral symmetry breaking scale) and are computed using
chiral perturbation theory.  They are typically of order a few
percent.}
\ENDTITLEPAGE

\eject

\noindent {\bf I.  Introduction}

Semileptonic $\dev$ and $\deva$ decays provide an interesting arena
to
test the validity of heavy quark symmetry.  They may also give a very
accurate determination of the element, $V_{cb}$, in the
Cabibbo-Kabayashi-Maskawa matrix.  Heavy quark spin-flavor symmetry
implies that as $m_b$ and  $m_c \rightarrow \infty$ the hadronic
matrix
elements needed for semileptonic $\dev$ and $\deva$ decay have the
form$^{[1]}$
$$	\eqalignno{{<D(v')|\bar c \gamma_\mu b|B(v)>\over \sqrt{m_B
m_D}} & = \beta (v + v')_\mu\,\, , &(1.1a)\cr
 	{<D^*(v',\epsilon) | \bar c \gamma_\mu \gamma_5 b| B(v)>\over
\sqrt{m_B m_{D^{*}}}} & = \beta ((1 + w)\epsilon_\mu^* - (\epsilon^*
\cdot v)
v'_\mu)\,\, , & (1.1b)\cr

	{<D^* (v', \epsilon) | \bar c \gamma_\mu b| B(v)>\over
\sqrt{m_B
m_{D^{*}}}} &= \beta i \epsilon_{\mu\nu\lambda\sigma} \epsilon^{*\nu}
v^{\prime\lambda} v^\sigma \,\, . & (1.1c)\cr}$$

In eqs. (1.1) $\beta$ depends on $w = v \cdot v'$ and it has a
calculable logarithmic dependence on the heavy $c$ and $b$ quark
masses
that arises from high momentum perturbative QCD effects.$^{[2,3,4]}$
Furthermore the value of $\beta$ is known at zero recoil, i.e., $w =
1$.$^{[1,5,6]}$  In the leading logarithmic approximation$^{[3,4]}$
(valid
for $m_b \gg m_c \gg \Lambda_{{\rm QCD}}$)
$$	\beta(1) = \left[{\alpha_s (m_b)\over \alpha_s
(m_c)}\right]^{-6/25} \,\, . \eqno (1.2)$$

Perturbative QCD corrections to eqs. (1.1) and (1.2), suppressed by
powers of
$\alpha_s(m_c)$ and $\alpha_s (m_b)$, are calculable and don't give
rise
to any loss of predictive power.$^{[2]}$  Since at zero recoil there
are
no $1/m_Q$ corrections to eqs. (1.1) and (1.2)$^{[7]}$ the exclusive
semileptonic
$\deva$ and $\dev$ decays may lead to a very precise determination of
$V_{cb}$.

Recently chiral perturbation theory has been used to examine, at zero
recoil, the power $(1/m_c)^{2+n}~ n = 0, 1,2,...$ corrections to the
matrix elements of eqs. (1.1).$^{[8]}$  It was found that these
corrections have a nonanalytic dependence on the light up and down
quark
masses of the form $\ell nm_\pi^2$ when $n = 0$ and $(1/m_\pi)^n$
when
$n = 1,2,3,...$~.  These corrections arise from one-loop Feynman
diagrams with virtual momenta small compared with the chiral symmetry
breaking scale.

In this paper we extend the work of Ref. [8] away from zero recoil.
The
corrections of order $(1/m_c)^{n+2}~n = 0,1,2,...$ that have a
nonanalytic dependence on the light up and down quark masses are
calculated for all $w$.

Unfortunately, away from zero recoil there are power corrections of
order $1/m_c$ and $1/m_b$ that are not calculable using chiral
perturbation theory.  These must be estimated using phenomenological
models (e.g., the nonrelativistic constituent quark model or QCD sum
rules) or lattice QCD methods.  Although the corrections of order
$1/m_c$ and $1/m_b$ are larger than those we can calculate it is
still interesting that some power corrections are calculable and it
is
important to verify that they are not anomalously large.

\noindent {\bf II.  Chiral Perturbation Theory for Semileptonic Form
Factors}

The ground state heavy mesons with $Q\bar q_a$ flavor quantum numbers
(Here $a = 1,2$ and $q_1 = u, q_2 = d$) have $s_\ell^{\pi_{\ell}} =
{1\over 2}^-$, for the spin parity of the light  degrees of freedom.
Combining the spin of the light degrees of freedom with the spin of
the
heavy quark gives (in the $m_Q \rightarrow \infty$ limit) two
degenerate
doublets consisting of spin zero and spin-one mesons  that are
denoted by

$P_a$ and  $P_a^*$ respectively.  In the case $Q = c, P_a = (D^0,
D^+)$ and $P_a^* = (D^{*0}, D^{*+})$ while for $Q = b, P_a = (B^-,

B^0$) and $P_a^* = (B^{*-}, B^{*0})$.  It is convenient to combine
the fields $P_a$ and $P_{a\mu}^*$ that destroy these mesons $(v^\mu
P_{a\mu}^* = 0)$ into a $4\times 4$ matrix $H_a$ given by

$$	H_a = \left({1 + \vslash\over 2}\right) (P_{a\mu}^*
\gamma^\mu -
P_a \gamma_5) \,\, . \eqno (2.1)$$
(This is a compressed notation.  In situations where the type of
heavy quark $Q$ and its four-velocity $v$ are important  the $4\times
4$
matrix is denoted by $H_a^{(Q)} (v)$).  It transforms under the heavy
quark spin symmetry group $SU(2)_v$ as
$$	H_a \rightarrow S H_a \,\, , \eqno (2.2)$$
where $S\epsilon SU(2)_v$ and under Lorentz transformations as
$$ 	H_a \rightarrow D(\Lambda) H_a D(\Lambda)^{-1} \,\, , \eqno
(2.3)$$
where $D(\Lambda)$ is an element of the $4\times 4$ matrix
representation of the Lorentz group.  It is also useful to introduce
$$	\eqalign{\bar H_a &= \gamma^0 H_a^{\dag} \gamma^0\cr
&= (P_{a\mu}^{*\dag} \gamma^\mu + P_a^{\dag} \gamma_5) {(1 +
\vslash)\over 2}

\,\, .\cr} \eqno (2.4)$$
For $\bar H_a$ the transformation laws corresponding to those in eqs.
(2.2) and (2.3) become $\bar H_a \rightarrow \bar H_a S^{-1}$ and
$\bar H_a
\rightarrow D(\Lambda) \bar H_a D(\Lambda)^{-1}$.

The strong interactions also have an approximate $SU(2)_L \times
SU(2)_R$ chiral symmetry that is spontaneously broken to the vector
$SU(2)_V$ isospin subgroup.   This
symmetry arises because the light up and down quarks have masses that
are small compared with the typical scale of the strong interactions.
(If

the strange quark is also treated as
light the chiral symmetry group becomes $SU(3)_L \times SU(3)_R$).
Associated with the spontaneous breaking of $SU(2)_L \times SU(2)_R$
chiral symmetry are the pions.  The low momentum strong interactions
of
these pseudo Goldstone bosons are described by a chiral Lagrangian
that
contains the most general interactions consistent with chiral
symmetry.
The effects of the up and down quark masses are included by adding
terms
that transform in the same way under chiral symmetry as the quark
mass
terms in the QCD Lagrangian.

The pions are incorporated in a $2\times 2$ unitary matrix
$$	 \Sigma = \exp \left({2iM\over f}\right) \eqno (2.5)$$
where
$$	M = \left[\matrix{\pi^0/\sqrt{2} & \pi^+\cr
\pi^- & -\pi^0/\sqrt{2}\cr}\right] \,\, , \eqno (2.6)$$
and $f \simeq 132$ MeV is the pion decay constant.  Under a chiral
$SU(2)_L \times SU(2)_R$ transformation
$$	\Sigma \rightarrow L\Sigma R^{\dag}\,\, , \eqno (2.7)$$
where $L\epsilon SU(2)_L$ and $R\epsilon SU(2)_R$.  It is convenient
when discussing the interactions of the $\pi$ mesons with the $P_a$
and
$P_a^*$ mesons to introduce
$$	\xi = \exp \left({iM\over f}\right) \,\, . \eqno (2.8)$$
Under a chiral $SU(2)_L \times SU(2)_R$ transformation
$$	\xi \rightarrow L\xi U^{\dag} = U\xi R^{\dag}\,\, , \eqno
(2.9)$$
where typically the special unitary matrix $U$ is a complicated
nonlinear function of $L, R$ and the pion fields.  However, for
transformations $V = L = R$ in the unbroken subgroup $U = V$.  We
assign
the heavy meson fields the transformation law
$$	H_a \rightarrow H_b U_{ba}^{\dag} \,\, , \eqno (2.10)$$
under chiral $SU(2)_L \times SU(2)_R$ (In eq. (2.10) and for the
remainder
of this paper repeated subscripts $a$ and $b$ are summed over $1,2$).

The low momentum strong interactions of pions with heavy $P_a$ and
$P_a^*$ mesons are described by the effective Lagrange
density$^{[9,10,11]}$
$$	{\cal L} = - i Tr \bar H_a v_\mu \partial^\mu H_a +  {1\over
2}
iTr \bar H_a H_b v^\mu (\xi^{\dag} \partial_\mu \xi + \xi
\partial_\mu
\xi^{\dag})_{ba}$$
$$	+ {1\over 2} ig Tr \bar H_a \gamma_\nu \gamma_5
H_b(\xi^{\dag}
\partial^{\nu} \xi - \xi \partial^\nu \xi^{\dag})_{ba} + ... \,\, ,
\eqno
(2.11)$$
where the ellipsis denote terms with more derivatives.  This Lagrange
density is the most general one invariant under $SU(2)_L \times
SU(2)_R$ chiral symmetry, heavy quark spin symmetry, parity and
Lorentz
transformations.  Heavy quark flavor symmetry implies that $g$ is
independent of the heavy quark mass.  Note that in eq. (2.11) factors
of
$\sqrt{m_P}$ and $\sqrt{m_{P^{*}}}$ have been absorbed into the $P_a$
and $P_a^{*\mu}$ fields so they have dimension ${3\over 2}$.

The coupling $g$ determines the $D^{*+} \rightarrow D^0 \pi^+$ decay
width
$$	\Gamma (D^{*+} \rightarrow D^0 \pi^+) = {1\over 6\pi} ~
{g^2\over f^2} |\vec p_\pi |^3\,\, . \eqno (2.12)$$
Using the measured branching ratio, for this decay$^{[12]}$ and the

limit on the $D^*$ width$^{[13]}$ gives the bound $g^2 \leq 0.5$.

It is possible to include the symmetry breaking effects of order
$m_q$
and $1/m_Q$ into the effective Lagrangian for pion heavy meson strong
interactions. In our calculations explicit chiral symmetry breaking
effects enter only through the nonzero pion mass.  Other chiral
symmetry
breaking effects are suppressed relative to the leading corrections
which we

calculate.

The $1/m_Q$ terms that break the spin-flavor heavy quark symmetry
give
rise to the additional terms
$$	\delta {\cal L}^{(2)} = {\lambda_2\over m_Q} Tr \bar H_a
\sigma^{\mu\nu} H_a \sigma_{\mu\nu} + {\lambda'_2\over m_Q} Tr \bar
H_a
H_a + ... \eqno (2.13)$$
in the Lagrange density.  The second term in eq. (2.13)
violates the heavy quark flavor symmetry but not the spin symmetry
and
the first term violates both the heavy quark spin and flavor
symmetries.  The ellipsis denote terms with derivatives.  Included in
these are, for example, the $1/m_Q$ correction to $g$.
The second term in eq. (2.13) can be removed by a spacetime dependent
phase

transformation on the heavy meson fields.  Therefore,
at  the leading order in chiral perturbation
theory, it is only the first term in eq. (2.13) that produces
violations of

heavy quark spin-flavor symmetry in the
low energy heavy meson Lagrangian. The
effect of  $\lambda_2$ is to shift the mass
of the pseudoscalar  relative to the  vector meson by

$$	\Delta^{(Q)} = m_{P^{*(Q)}} - m_{P^{(Q)}} \,\, , \eqno
(2.14)$$
which distinguishes the heavy meson propagators.
Explicitly, $\Delta^{(Q)} = - 8 \lambda_2/m_Q$,
which determines $\lambda_2 \approx (170 {\rm MeV})^2$.

 Heavy quark mass suppressed
operators in which the pion couples can be neglected at leading
order in chiral perturbation theory.

Lorentz and parity invariance of the strong interactions implies that
the hadronic matrix elements needed for semileptonic $\dev$ and
$\deva$
decay have the form
$$	\eqalignno{{<D(v')|\bar c \gamma_\mu b|B(v)>\over \sqrt{m_B
m_D}} &= \tilde{f}_+ (v + v')_\mu + \tilde{f}_- (v - v')_\mu &
(2.15a)\cr
	{<D^*(v',\epsilon) | \bar c \gamma_\mu \gamma_5 b|B(v)>\over
\sqrt{m_B m_{D^{*}}}} &= \tilde{f} \epsilon_\mu^* + \tilde{a}_+
(\epsilon^*
\cdot v) (v + v')_\mu &(2.15b)\cr
	& + \tilde{a}_- (\epsilon^* \cdot v) (v - v')_\mu\cr
	{<D^* (v',\epsilon)|\bar c \gamma_\mu b|B(v)>\over \sqrt{m_B
m_{D^{*}}}} & = i \tilde{g} \epsilon_{\mu\nu\lambda\sigma}
\epsilon^{*\nu} v^{\prime\lambda} v^\sigma & (2.15c)\cr}$$

The operator $\bar c \Gamma b$ is a singlet under $SU(2)_L \times
SU(2)_R$ and in chiral perturbation theory its $B(v) \rightarrow
D(v')$
and $B(v) \rightarrow D^* (v')$ matrix are given by those of
$$	\bar c \Gamma b  = - \beta (w) Tr \bar H_a^{(c)} (v') \Gamma
H_a^{(b)} (v) + ... \,\, , \eqno (2.16)$$
where the ellipsis denote terms with derivatives, factors of $m_q$
and
factors of $1/m_Q$.  Evaluating at tree level $B\rightarrow D$ and $B
\rightarrow D^*$ matrix elements of the rhs of eq. (2.16) for $\Gamma
=
\gamma_\mu$ and $\Gamma = \gamma_\mu \gamma_5$ gives eqs. (1.1).
This implies the relations
$$	\tilde{f}_+ = \beta \,\, , \qquad \tilde{f}_- = 0 \,\, ,$$
$$	\tilde{a}_+ + \tilde{a}_- = 0 \,\, , \qquad \tilde{a}_+ -
\tilde{a}_- = - \beta \,\, ,$$
$$	\tilde{f} = (1 + w) \beta \,\, , \qquad \tilde{g} = \beta
\,\, ,
\eqno (2.17)$$
for the form factors in eqs. (2.15).

The leading corrections to eqs. (2.17) that have a nonanalytic
dependence on the
light quark masses arise from the one loop Feynman diagrams in Fig. 1
and wave function renormalization.  In Fig. 1 the shaded square
denotes
a vertex from eq. (2.16) and a dot is a $P^* P\pi$ or $P^* P^* \pi$
vertex proportional to $g$ from the chiral Lagrange density in eq.
(2.11). Explicit calculation of these Feynman diagrams gives that
their
contribution to the form factors is
$$	\eqalignno{\delta \tilde{f}_+ &= - {3ig^2\beta\over f^2}
\bigg\{[w+2] I_1 (\Delta, w) + [w^2 - 1] I_2 (\Delta, w)\cr
	& \qquad \qquad - {3\over 2} I_3 (\Delta, w) - {3\over 2} I_3
(0, w)\bigg\} & (2.18a)\cr
\delta \tilde{f}_- &= 0 & (2.18b)\cr
\delta \tilde{f} &= - {3ig^2\beta\over f^2} [w+1] \bigg\{I_1
(-\Delta, w) - {1\over 2} I_3 (- \Delta, w)\cr
	& \qquad \qquad + [w+1] I_1 (0,w) + [w^2 -1] I_2 (0,w)\cr
	& \qquad \qquad - {5\over 2} I_3 (0,w) \bigg\} & (2.18c)\cr
\delta(\tilde{a}_+ - \tilde{a}_-) &= {3ig^2\beta\over f^2} \bigg\{ -
[w+1] I_2 (-\Delta, w) - {1\over 2} I_3 (-\Delta, w)\cr
	& \qquad \qquad + [w + 2] I_1 (0,w) + [w^2 + w] I_2 (0,w)\cr
	& \qquad \qquad - {5\over 2} I_3 (0,w) \bigg\} & (2.18d)\cr
\delta (\tilde{a}_+ + \tilde{a}_-) &= - {3ig^2\beta\over f^2} \Bigg\{
-
I_1 (- \Delta, w) - [w+1] I_2 (- \Delta, w)\cr
	& \qquad \qquad + I_1 (0,w) + [w + 1] I_2 (0,w) \Bigg\} &
(2.18e)\cr
\delta \tilde{g} &= - {3ig^2\beta\over f^2} \Bigg\{ I_1 (- \Delta, w)
-
{1\over 2} I_3 (- \Delta, w)\cr
	& \qquad \qquad + [w+1] I_1 (0,w) + [w^2 - 1] I_2 (0,w) \cr
	& \qquad \qquad - {5\over 2} I_3 (0,w)\Bigg\}\,\, . &
(2.18f)\cr}$$
In Eqs. (2.18) the integrals $I_1, I_2$ and $I_3$ are
$$	\eqalignno{I_1 &= \mu^{4-n} \int_0^\infty d\alpha
\int_0^\infty
d\beta \int {d^n k\over (2\pi)^n} ~ {k^2\over [ k^2 - (\alpha^2 +
\beta^2
+ 2\alpha\beta w + 2 \Delta \alpha + m_\pi^2) +i\epsilon]^3}~,~~ &
(2.19)\cr
	I_2 &= 4\mu^{4-n} \int_0^\infty d\alpha \int_0^\infty d\beta
\int {d^nk\over (2\pi)^n} ~ {\alpha\beta\over [k^2 - (\alpha^2 +
\beta^2 +
2\alpha\beta w + 2 \Delta \alpha + m_\pi^2)+i\epsilon]^3}~,~~~~~~~ &
(2.20)\cr
	I_3 &= \mu^{(4-n)} \int_0^\infty d\alpha \int {d^nk\over
(2\pi)^n} ~ {\alpha k^2\over [k^2 - (\alpha^2 + 2\Delta\alpha +
m_\pi^2)+i\epsilon]^3}\,\, , & (2.21)\cr}$$
where $\mu$ is the subtraction point and $\Delta = m_{D^{*}} - m_D$.
For simplicity we have taken $m_b \rightarrow \infty$, so that
corrections suppressed by factors of $(1/m_b)$ are neglected.  Note
that
$\tilde{f}_-$ receives no corrections and $\delta\tilde{f}/\tilde{f}
=
\delta \tilde{g}/\tilde{g}, \delta \tilde{a}_+ = - \delta
\tilde{g}/2$.

In calculating the Feynman diagrams the amplitudes were reduced to
scalar integrals using four-dimensional algebra (e.g., $Tr \gamma_\mu
\gamma_\nu = 4g_{\mu\nu}, k_\mu k_\nu \rightarrow {1\over 4} k^2
g_{\mu\nu}$, etc).  The resulting scalar integrals were then
continued
to $n$ dimensions.  This regularization procedure is similar to that
used in supersymmetry.  It is easy to see that it preserves the heavy
quark spin-flavor symmetry when $\Delta = 0$.  Then the corrections
in
eqs. (2.18) can be absorbed into the redefinition, $\beta \rightarrow
\beta + \delta \beta$, where
$$	\delta\beta = - {3ig^2\over f^2} \beta \left\{(w+2) I_1 (0,w)
+
(w^2 - 1) I_2 (0,w) - 3 I_3 (0,w)\right\} \,\, . \eqno (2.22)$$
Furthermore, since $I_3 (0,w) = I_1 (0,w), \delta\beta$ vanishes at
zero
recoil.

In eqs. (2.18) the terms containing $I_3$ result from wave function
renormalization.  The integrations over $\alpha$ and $\beta$ arise
from
combining denominators using the trick introduced in Ref. [2].

The integrals have the form

$$	I_j (\Delta, w) = {i\over 16\pi^2} [m_\pi \Delta E_j (w) +
\Delta^2 \ell n (m_\pi^2/\mu^2) G_j (w)$$
$$	+ \Delta^2 F_j (\Delta/m_\pi, w)] + ... \,\, , \eqno (2.23)$$
where $F_j(0,w) = 0$ and the ellipsis denotes terms that are
independent
of $\Delta$ or have an analytic dependence on the light up and down
quark
masses.  Evaluating the integrals gives
$$	\eqalignno{E_1 (w) &= \pi/(w+1) \,\, , &(2.24a)\cr
E_2 (w) &= - \pi/(w+1)^2 \,\, , & (2.24b)\cr
E_3 (w) &= \pi \,\, , & (2.24c)\cr
G_1 (w) &= - {1\over 2(w^2 - 1)} [w - r(w)] \,\, & (2.25a)\cr
G_2 (w) &= {1\over 2(w^2 - 1)^2} [w^2 + 2 - 3wr(w)] \,\, , &
(2.25b)\cr
G_3 (w) &= - 1 \,\, , & (2.25c)\cr}$$
where
$$	r(w) = {1\over \sqrt{w^2 - 1}} \ell n (w + \sqrt{w^2 - 1})
\,\,
. \eqno (2.26)$$

For $F_{1,2}$ the expressions are more complicated and we leave them
as
one-dimensional integrals,
$$	F_1 (x,w) = {1\over x^2} \int_0^{\pi/2} d\theta {a\over (1 +
w\sin 2\theta)} \bigg\{\pi \left(\sqrt{1 - a^2} - 1\right)$$
$$	- 2 \left( \sqrt{1 - a^2} ~arc\tan~ \left({a\over\sqrt{1 -
a^2}}\right) - a\right)\Bigg\} \,\, , \eqno (2.27a)$$
$$	F_2 (x,w) = {1\over x^2} \int_0^{\pi/2} d\theta {a\sin
2\theta\over (1+w \sin 2\theta)^2}  \Bigg\{- {3\pi\over 2}
\left(\sqrt{1-a^2} - 1\right)$$
$$	+ {\pi a^2\over 2\sqrt{1 - a^2}} + \left(\left[{3 - 4
a^2\over
\sqrt{1 - a^2}}\right] ~arc\tan~ \left({a\over \sqrt{1 -a^2}}\right)
-
3a \right) \Bigg\} \,\, , \eqno (2.27b)$$
where
$$	a = {x \cos \theta\over\sqrt{1 + w\sin 2\theta}} \,\, .\eqno
(2.28)$$
Figs. 2,3,4 and 5 present plots of $F_1 (1,\omega), F_1 (-1, \omega),

F_2(1,\omega)$ and $F_2 (-1,w)$.  Finally
$$	F_3 (x,w) = {1\over x} \Bigg\{ \pi \left( \sqrt{1 - x^2} - 1
\right) - 2 \left(\sqrt{1 - x^2} ~arc\tan~ \left({x\over \sqrt{1 -
x^2}}\right) - x \right) \Bigg\} \,\, . \eqno (2.29)$$

At $x = \pm 1$ the above becomes $F_3(1,w) = (2 - \pi)$ and
$F_3(-1,w) =
(2 + \pi)$.
Combining these results gives the correction to the form factors that
have a non-analytic dependence on the up and down quark masses.   For
the
corrections of order $(1/m_c)^{n+2}, n = 0,1,2, ...$ we have
$$	\delta f_+ = {3g^2 \beta\over (4\pi f)^2} \Delta^2
\Bigg\{{1\over 2} \left({3w + 1 - 2r\over w + 1}\right) \ell n
(m_\pi^2/
\mu^2)$$
$$	+ (w+2) F_1 (\Delta/m_\pi, w) + (w^2 - 1) F_2
(\Delta/m_\pi,w) -
{3\over 2} F_3 (\Delta/m_\pi, w) +... \Bigg\} \eqno (2.30a)$$
$$	\delta \tilde{f} = {3g^2\beta\over (4\pi f)^2} \Delta^2 (w +
1)
\Bigg\{{1\over 2} \left(1 - \left({w-r\over w^2-1}\right)\right) \ell
n
(m_\pi^2/\mu^2)$$
$$	F_1 (-\Delta/m_\pi,w) - {1\over 2} F_3 (-\Delta/m_\pi, w)
+...
\Bigg\} \eqno (2.30b)$$
$$	\delta(\tilde{a}_+ - \tilde{a}_-) = - {3g^2\beta^2\over (4\pi
f)^2} \Delta^2 \Bigg\{{1\over 2} \left( 1 - {(w+1) (w^2 - 3wr+2)\over
(w^2 - 1)^2}\right) \ell n (m_\pi^2/\mu^2)$$
$$	- (w+1) F_2 (- \Delta/m_\pi, w) - {1\over 2} F_3
(-\Delta/m_\pi,w) + ...\Bigg\} \eqno (2.30c)$$
$$	\delta (\tilde{a}_+ + \tilde{a}_-) = {3g^2\beta^2
\Delta^2\over
(4\pi f)^2} \Bigg\{{1\over 2}\left({r(2w^2 + 3w + 1) - (w^2 + 3w +
2)\over

(w^2 - 1)^2}\right) \ell n (m_\pi^2/\mu^2)$$
$$	- F_1 (-\Delta/m_\pi, w) - (w + 1) F_2 (-\Delta/m_\pi, w)
+...
{}~\Bigg\} \eqno (2.30d)$$
and
$$	\delta\tilde{f}_- = 0, \quad \delta\tilde{g} =
\delta\tilde{f}/(w+ 1) \eqno (2.30e)$$

Eqs. (2.30) are the main results of this paper.  In these equations
the
ellipsis denote terms that are less singular as the light quark
masses
go to zero.  The subtraction point dependence of these terms cancels
that in the logarithm.  Terms independent of $\Delta$ have been
absorbed into
a redefinition of $\beta$.

We have also computed the corrections of order $(1/m_c)$ that depend
on
the light up and down quark masses as $m_q^{1/2}$.  They are
$$	\delta\tilde{f}_+ = - {3g^2\beta m_\pi \Delta\over 32\pi f^2}
\left({3(w - 1)\over w + 1}\right) \eqno (2.31a)$$
$$	\delta\tilde{f} = {3g^2 \beta m_\pi \Delta\over 32 \pi f^2}
(w -
1) \eqno (2.31b)$$
$$	\delta (\tilde{a}_+ - \tilde{a}_-) = - {3g^2\beta m_\pi
\Delta\over 32\pi f^2} \left({w-1\over w+1}\right) \eqno (2.31c)$$
$$	\delta (\tilde{a}_+ + \tilde{a}_-) = 0 \eqno (2.31d)$$
and
$$	\delta\tilde{f}_- = 0, \quad \delta\tilde{g} =
\delta\tilde{f}/(w+1) \eqno (2.31e)$$
These corrections are less important than the $1/m_c$ corrections
that
are independent of $m_q$.  Note that eqs. (2.31a) and (2.31b) are
consistent with Luke's theorem since they vanish at $w = 1$.  In fact
at
order $1/m_c$ the $m_q^{1/2}$ corrections to all the form factors
vanish
at $w = 1$.

Experimentally $\Delta$ is larger than $m_\pi$.  The $B \rightarrow
D^*$ form
factors have imaginary parts but because $\Delta$ is close to
$m_\pi$ they are negligible.  It is a very good approximation to
evaluate the functions $F_j$ at $\Delta/m_\pi = 1$.
\vfil\eject
\noindent {\bf III.  Discussion}

In the limit $m_c, m_b \rightarrow \infty$ heavy quark symmetry
implies
that the form factors for $B(v) \rightarrow D(v')e\bar\nu_e$ and
$B(v)
\rightarrow D^* (v') e\bar\nu_e$ decay can be expressed in terms of a
single universal function of $w = v \cdot v'$ and that the value of
this
function at $w = 1$ is known.  In this paper we examined, using
chiral
perturbation theory, corrections to the heavy quark symmetry
relations that arise from the non-zero value of the charm quark mass.
We calculated corrections of order $(1/m_c)^{n+2}~ n = 0,1,2,...$
that go
as $\ell n m_\pi^2$ when $n = 0$ and as $(1/m_\pi)^n$ when $n = 1,2,3
...$~.  These arise from physics well below the chiral symmetry
breaking
scale.  The factors of $1/m_c$ occur from insertions of the $D^* - D$
mass difference $\Delta$.  In our calculations other sources of heavy
quark

symmetry breaking, are less important for

very small up and down quark masses.

Since the value of $\Delta$ is close to $m_\pi$ all the corrections
of
order $(1/m_c)^{n + 2}~ n = 1,2,3,...$ are equally important.  In
eqs.
(2.30) these corrections occur in the functions $F_j$.  Our
calculation
of the effects of this order is enhanced by a factor of $(1/m_\pi)$
over
terms we neglected.

At order $1/m_c^2$ the terms we calculated are only enhanced by $\ell
n
m_\pi^2$ over those we neglected.  The pion mass is not small enough
to
have complete confidence in our calculation of the $1/m_c^2$
corrections
to the semileptonic decay form factors.  For example, other effects
that don't arise from the operator $\bar h_v^{(c)} g_s\sigma^{\mu\nu}
T^A h_v^{(c)} G_{\mu\nu}^A$ may be more important.

At zero recoil there are no $1/m_c$ or $1/m_b$ corrections and Ref.
[8]
used chiral perturbation theory to compute the corrections to
$\tilde{f}_+$ and $\tilde{f}$ of order $(1/m_c)^{n + 2} ~n =
0,1,2,...$~.

At $w = 1$ our expressions for these
form factors agree with those in Ref. [8].  This paper contains the
extension of the results of Ref. [8] away from zero recoil.

We also computed the order $1/m_c$ corrections that have a
non-analytic
dependence on the light quark masses of the form $m_q^{1/2}$.
However,
the dominant corrections (away from zero recoil) are those of order
$1/m_c$ (and $1/m_b$) that are independent of the light quark masses.
These are not calculable using chiral perturbation theory and must be
estimated using phenomenological models$^{[14]}$ or lattice Monte
Carlo
methods.

For $g^2 = 0.5$ (the present experimental limit) the corrections we
calculated are typically a few percent.  The constituent quark model
suggests that $g$ is around unity but it is certainly possible that
it
is much smaller.$^{[15]}$

We have used chiral $SU(2)_L \times SU(2)_R$ so the effects discussed
in
this paper are associated with small virtual momenta of order the
pion
mass.  It is possible to extend the calculations to chiral $SU(3)_L
\times SU(3)_R$, however, the kaon mass is too large to have
confidence
in such calculations.$^{[16]}$  For example, at momentum scales
around
the kaon mass it is difficult to justify the neglect of contributions
from excited heavy mesons$^{[17]}$ in the loop of Fig. 1.

\noindent {\bf References}

\item{1.}  N. Isgur and M.B. Wise, Phys. Lett., {\bf B232} 113
(1989);
Phys. Lett., {\bf B237} 527 (1990).

\item{2.}  A. Falk, et al., Nucl. Phys., {\bf B343} 1 (1990).

\item{3.}  M.B. Voloshin and M.A. Shifman, Sov. J. Nucl. Phys., {\bf
45}
292 (1987).

\item{4.}  H.D. Politzer and M.B. Wise, Phys. Lett., {\bf B206} 681
(1988); {\bf B208} 504 (1988).

\item{5.}  S. Nussinov and W. Wetzel, Phys. Rev., {\bf D36} 130
(1987).

\item{6.}  M.B. Voloshin and M.A. Shifman, So. J. Nucl. Phys., {\bf
47}
199 (1988).

\item{7.}  M. Luke, Phys. Lett., {\bf B252} 447 (1990).

\item{8.}  L. Randall and M.B. Wise, Phys. Lett., {\bf B303} 139
(1993).

\item{9.}  M.B. Wise, Phys. Rev., {\bf D45} 2188 (1992).

\item{10.}  G. Burdman and J. Donoghue, Phys. Lett., {\bf B208} 287
(1992).

\item{11.}  T.M. Yan, et al., Phys. Rev., {\bf D46} 1148 (1992).

\item{12.}  The CLEO Collaboration, (F. Butler, et al.), Phys. Rev.
Lett., {\bf 69} 2041 (1992).

\item{13.}  The ACCMOR Collaboration (S. Barlag, et al.), Phys.
Lett.,
{\bf B278} 480 (1992).

\item{14.}  A.F. Falk and M. Neubert, SLAC-PUB-5897 (1992)
unpublished.

\item{15.}  W.A. Bardeen and C.T. Hill, SSCL-PP-243 (1993)
unpublished.

\item{16.}  L. Randall and E. Sather, Phys. Lett., {\bf B303} 345
(1993).

\item{17.}  A.F. Falk, SLAC-PUB-6055 (1993) unpublished.

\noindent {\bf Figure Captions}

\item{~~~~~Fig. 1.}  Feynman diagrams that give corrections to form
factors
$\tilde{f}_\pm, \tilde{f}, \tilde{a}_\pm, \tilde{g}$.

\item{~~~~~Fig. 2.}  Plot of $F_1(1,w)$ versus $w$.

\item{~~~~~Fig. 3.}  Plot of $F_1(-1,w)$ versus $w$.

\item{~~~~~Fig. 4.}  Plot of $F_2(1,w)$ versus $w$.

\item{~~~~~Fig. 5.}  Plot of $F_2(-1,w)$ versus $w$.

\bye